\documentclass{jfm_old}

\usepackage{graphicx}
\usepackage{newtxtext}
\usepackage{newtxmath}
\usepackage{amsmath}
\usepackage{natbib}
\usepackage{hyperref}
\hypersetup{
    colorlinks = true,
    urlcolor   = blue,
    citecolor  = black,
}

\newcommand{\RomanNumeralCaps}[1]
\linenumbers

\usepackage{bm}

\newcommand{\inv}{\mathcal{I}}
\newcommand{\Tr}{\textrm{Tr}}

\newcommand{\avg}[1]{\left\langle #1 \right\rangle}

\newcommand{\vect}[1]{\bm{#1}}
\newcommand{\tens}[1]{\mathsfbi{#1}}
\newcommand{\fvel}{f}

\title{Are there higher-order Betchov homogeneity constraints for incompressible isotropic turbulence?}

\author{
Maurizio Carbone\aff{1}
\corresp{\email{maurizio.carbone@ds.mpg.de}}
\and
Michael Wilczek\aff{1,2}
}

\affiliation{\aff{1}Max Planck Institute for Dynamics and Self-Organization, Am Fa{\ss}berg 17, 37077 Göttingen, Germany
\aff{2} Theoretical Physics I,  University of Bayreuth, Universit\"atsstr.\ 30, 95447 Bayreuth, Germany}

\begin{document}
\maketitle

\begin{abstract}
Incompressible and statistically homogeneous flows obey exact kinematic relations. The Betchov homogeneity constraints
(Betchov, \textit{J.~Fluid Mech.}, vol.~1, 1956, pp.~497--504)
for the average principal invariants of the velocity gradient are among the most well known and extensively employed homogeneity relations. These homogeneity relations have far-reaching implications for the coupled dynamics of strain and vorticity, as well as for the turbulent energy cascade.
Whether the Betchov homogeneity constraints are the only possible ones or whether additional homogeneity relations exist has not been proven yet. Here we show that the Betchov homogeneity constraints are the only homogeneity constraints for incompressible and statistically isotropic velocity gradient fields.
We also extend our results to derive homogeneity relations involving the velocity gradient and other dynamically relevant quantities, such as the pressure Hessian.
\end{abstract}

\begin{keywords}
homogeneous isotropic turbulence, statistical turbulence theory
\end{keywords}

\section{Introduction}

The velocity gradients, i.e.~spatial derivatives of the velocity field, $\tens{A} = \bnabla\vect{ u}$, contain a wealth of information about small-scale turbulence, including the topology of vorticity and strain \citep{Meneveau2011}.
The moments of the velocity gradients of an incompressible and statistically homogeneous field obey exact kinematic relations \citep{Betchov1956}.
The two so-called Betchov constraints for the velocity gradient principal invariants, namely the matrix traces $\Tr(\tens{A}^2)$ and $\Tr(\tens{A}^3)$, are of central importance for a statistical description of the turbulent dynamics \citep{Davidson2004}.
The first Betchov constraint states that the second principal invariant of the velocity gradient is on average zero, $\avg{\Tr(\tens{A}^2)}=0$, which implies the proportionality between the mean dissipation rate and the mean squared vorticity,
\begin{align}
\varepsilon = \nu\avg{ \vect{\omega}^2}.
\label{first_Betchov}
\end{align}
Here $\nu$ is the kinematic viscosity of the fluid, $\varepsilon = 2\nu\avg{\Tr\big(\tens{S}^2\big)}$ is the mean dissipation rate, $\tens{S}$ is the strain rate and $\vect{\omega}=\bnabla\times\vect{u}$ is the vorticity.
The second Betchov relation $\avg{\Tr(\tens{A}^3)}=0$, connects strain self-amplification and vortex stretching,
\begin{align}
4\big\langle\Tr\big(\tens{S}^3\big)\big\rangle =
-3\big\langle\vect{\omega}\bcdot\left(\tens{S}\vect{\omega}\right)\big\rangle.
\label{second_Betchov}
\end{align}
While equation \eqref{first_Betchov} constrains the strain-rate and vorticity magnitudes, relation \eqref{second_Betchov} constrains their production rates.
The latter relation was derived first by \cite{Townsend1951} and then rederived and extensively used  by \cite{Betchov1956}. It allows to characterize the average turbulent energy cascade in physical space \citep{Davidson2004,Carbone2020,Johnson2020}
and to predict the preferential configuration of the strain-rate eigenvalues
\citep{Betchov1956}.
It also implies that the vortex stretching has a positive average in presence of an average forward energy cascade, related to the negative skewness of the longitudinal velocity increment and gradient statistics. This positive average has implications, for example, on the vorticity magnitude and orientation relative to the strain rate \citep{Tsinober2009,Tom2021} and on the attenuation of extreme velocity gradients \citep{Buaria2020}.
Additionally, relations (\ref{first_Betchov}, \ref{second_Betchov}) have their analogues for the velocity structure functions  \citep{Hill1997}.
Those relations for the velocity structure functions are related to the ones for the velocity gradients through a simple Taylor expansion at small scales, and at larger scales through a filtered velocity gradient corrected for compressible effects \citep{Carbone2020}.

Applications of the homogeneity relations (\ref{first_Betchov}, \ref{second_Betchov}) are not limited to the theoretical understanding of turbulence, but they carry over to the modelling of turbulent flows.
For example, stochastic models for the velocity gradient should in principle obey the constraints \eqref{first_Betchov} and \eqref{second_Betchov} \citep{Johnson2016} which help to reduce the number of free parameters in such models \citep{Leppin2020}. The homogeneity relations can also be used to improve the performance of neural networks designed for machine learning of turbulent flows by including them into the training \citep{Tian2021,Momenifar2021}.

The Betchov relations (\ref{first_Betchov}, \ref{second_Betchov}) follow by writing the matrix traces $\Tr(\tens{A}^2)$ and $\Tr(\tens{A}^3)$ as the divergence of a vector field.
Then, because of statistical homogeneity, the average of such traces is zero since a spatial derivative can be factored out, and it acts on an average which does not depend on space explicitly.
For example, due to incompressibility, the second principal invariant of the velocity gradient can be rewritten as
\begin{align}
\Tr\big(\tens{A}^2\big) =
\nabla_j u_i \nabla_i u_j =
\nabla_i\left(u_j \nabla_j u_i\right),
\label{example_first_Betchov}
\end{align}
so that its average vanishes, and relation \eqref{first_Betchov} follows.
An analogous procedure applies to retrieve equation \eqref{second_Betchov}, since the third principal invariant can be expressed as
\begin{align}
\Tr\big(\tens{A}^3\big) =
\nabla_j u_i \nabla_k u_j \nabla_i u_k =
\nabla_i\left(u_j\nabla_k u_i\nabla_j u_k - \frac{1}{2} u_i \nabla_k u_j\nabla_j u_k \right).
\label{example_second_Betchov}
\end{align}
However, while it is straightforward to check the validity of the Betchov homogeneity relations for the velocity gradient, it is more complicated to show whether those homogeneity relations are the only possible ones or if additional constraints exist.
If there existed higher-order constraints we could, for example, improve the current reduced-order models of the velocity gradient dynamics just by imposing these additional homogeneity constraints.

Most of the previous attempts to find higher-order homogeneity relations were based on swapping the spatial derivatives: a scalar contraction of powers of the velocity gradient is manipulated by factoring out the spatial derivative, in order to rewrite the contraction as the divergence of some quantity (if possible). Attempts to obtain relations for the fourth-order moments of the velocity increments/gradients through this derivative-swapping procedure include \cite{Hill1997,Hierro2003,Bragg2021}.
However, it is very difficult
to show the completeness of the homogeneity constraints for the incompressible gradient through this derivative-swapping approach. One would need to consider linear combinations of infinitely many contractions of the velocity gradients and try to recast them into the spatial derivative of some field.
In this framework, \cite{Siggia1981} showed that no homogeneity constraints exist on polynomials of fourth-order velocity gradient invariants.

The scenario is analogous to the search for inviscid invariants of the Navier-Stokes equations \citep{Majda2001}, which are central for the occurrence of cascades \citep{Alexakis2018}.
While it is straightforward to check the conservation of kinetic energy  and helicity in the incompressible three-dimensional Euler equations by a derivative-swapping procedure \citep{Majda2001}, it is much more involved to show whether  those conserved quantities are the only possible ones or if additional ones exist.
This completeness question has been answered by \cite{Serre1984} for the incompressible Euler equations and by \cite{Enciso2016} for volume-preserving diffeomorphisms.

In this work, we investigate the existence of higher-order homogeneity constraints for the incompressible velocity gradient using tensor function representation theory
\citep{Zheng1994,Itskov2015}.
The analysis  allows identifying the homogeneity relations as the solutions of a system of partial differential equations, and it shows that no additional homogeneity constraints for the incompressible velocity gradient exist other than the ones already known from \cite{Betchov1956}.

\section{An equation encoding  the homogeneity constraints on the velocity gradient}

We consider a three-dimensional, incompressible and statistically homogeneous and isotropic velocity field $\vect{u}(\vect{x},t)$ together with its spatial gradient $\tens{A}=\bnabla \vect{u}$ ($A_{ij}=\nabla_j u_i$ in Cartesian  component notation).
We search for homogeneity relations for scalar single-point statistics of the velocity gradient only.

Incompressbility implies that $\Tr(\tens{A})=0$, while homogeneity implies translational invariance of ensemble averages, i.e.~it renders them independent of the spatial coordinate $\vect{x}$.
As a consequence, any scalar field $\phi$ that is the divergence of a vector field, $\phi=\bnabla\bcdot \vect{F}$, has zero ensemble/spatial average
\begin{equation}
\avg{\phi} = \avg{\bnabla\bcdot \vect{F}} = \bnabla\bcdot\avg{\vect{F}} = 0.
\label{hom}
\end{equation}
For example, 
in the first Betchov relation $\avg{\mathrm{Tr}\left(\tens A^2\right)}=0$, 
the vector field is $\vect{F}=\tens{A}\vect{u}$ ($F_i=A_{ij}u_j$ in component notation, see \eqref{example_first_Betchov}).
To generalize this, we search for scalar functions of the velocity gradient $\phi(\tens{A})$ which are the divergence of a vector field $\vect{F}$.

The vector field $\vect{F}$ is in general a \textit{functional} of the velocity field $\vect{u}(\vect{x},t)$.
We restrict the analysis to \textit{functions} of the velocity and its spatial derivatives, $\vect{F}(\vect{u},\tens{A},\bnabla \tens{A},\bnabla (\bnabla \tens{A}),\dots)$, because we search for homogeneity relations on the single-point statistics of the velocity gradient.
By restricting the analysis to functions of the velocity and its spatial derivatives, we are implicitly assuming isotropy.
Indeed, in a statistically isotropic flow, the governing equations and associated boundary conditions do not introduce any characteristic direction. Therefore, in that statistically isotropic situation, the velocity and velocity gradients are all the possible variables upon which the vector $\vect{F}$ can depend. We thereby exclude, for example, rotations of the frame of the flow, anisotropic forcing, boundary layers, etc.

Focusing on single-point statistics of isotropic flows we have $\vect{F} = \vect{F}(\vect{u},\tens{A},\bnabla \tens{A},\bnabla (\bnabla \tens{A}),\dots)$, so that the corresponding $\phi(\tens{A})$ is, by chain rule and in component notation,
\begin{align}
\phi(\tens{A}) = 
\nabla_i F_i \left(\vect{u}(\vect{x},t), \dots\right) =
\frac{\partial F_i}{\partial u_{p}} A_{pi} +  
\frac{\partial F_i}{\partial A_{pq}}\nabla_i A_{pq} +   
\frac{\partial F_i}{\partial \left(\nabla_k A_{pq}\right)} \nabla_i\left(\nabla_k A_{pq}\right) + \dots
\label{phi_general}
\end{align}
The fact that the left-hand side of equation \eqref{phi_general} depends only on the velocity gradient strongly constrains the functional form of the vector field $\vect{F}$.
Namely, the right-hand side of equation \eqref{phi_general} should explicitly involve neither the velocity $\vect{u}$ nor the gradients of the velocity gradient, $\bnabla \tens{A}$, $\bnabla (\bnabla \tens{A})$, etc.
This implies that all the terms on the right-hand side of \eqref{phi_general} featuring gradients of the velocity gradient should identically cancel, while only $\partial F_i/\partial u_{p} A_{pi}$ can contribute to $\phi$.
Moreover, the part of $\partial F_i/\partial u_{p} A_{pi}$ that contributes to $\phi$ can depend only on $\tens{A}$.
Therefore, we just need to consider vector functions of the velocity and velocity gradient, $\vect{F}(\vect{u},\tens{A})$, that are linear in the velocity.
Based on this,
equation \eqref{phi_general} splits into
\begin{subequations}
\begin{align}
&\phi(\tens{A}) = 
\frac{\partial F_i}{\partial u_{p}}\left(\vect{u},\tens{A}\right)A_{pi},
\label{phi_u}
\\
&\frac{\partial F_i}{\partial A_{pq}}\left(\vect{u},\tens{A}\right)\nabla_i A_{pq} = 0.
\label{phi_A}
\end{align}
\label{phi}
\end{subequations}
Equation 
\eqref{phi_A} yields the main differential equation to determine $\vect{F}$.
The gradient of the gradient, $\nabla_iA_{pq}=\nabla_i\nabla_q u_p$, is symmetric in $i,q$, so that only the part of $\partial F_i/\partial A_{pq}$ that is symmetric in $i,q$ contributes to \eqref{phi_A}.
Additionally, the contractions $i,p$ and $q,p$ of $\nabla_iA_{pq}$ are zero by incompressiblity.
Therefore, $\vect{F}$ solves equation \eqref{phi_A} only if, for some vector $\vect{v}$,
\begin{align}
\frac{\partial F_i}{\partial A_{pq}} + \frac{\partial F_q}{\partial A_{pi}} = v_i\delta_{pq} + v_q\delta_{pi}.
\label{A_constr}
\end{align}
Here,  $\delta_{ij}$ denotes the Kronecker delta, and the vector $\vect{v}$ is easily determined by contracting two of the free indices, 
e.g.~$v_p=\partial F_{k}/\partial A_{pk}$.

Equations (\ref{phi}, \ref{A_constr}) allow making the search for homogeneity constraints more systematic: instead of attempting to factor out the spatial derivatives in tensor contractions of velocity gradients, we need to solve a system of partial differential equations.
Solving equation \eqref{A_constr} for vectors $\vect{F}$ that are linear in $\vect{u}$ yields all possible vectors $\vect{F}(\vect{u},\tens{A})$, whose divergence depends only on the velocity gradient, $\phi(\tens{A}) = \bnabla\bcdot \vect{F}(\vect{u},\tens{A})$, as in \eqref{phi_u}.
Therefore, finding all solutions of \eqref{A_constr} that are linear in the velocity amounts to deriving all possible homogeneity constraints on scalar functions of an incompressible and statistically isotropic velocity gradient.

\section{Tensor function representation of the homogeneity constraints}

In the following we construct the general isotropic tensor function $\vect{F}(\vect{u},\tens{A})$.
Tensor function representation theory \citep{Weyl1946,Rivlin1955,Pennisi1987,Zheng1994,Itskov2015} allows writing all possible vector functions $\vect{F}$ of the generating vector $\vect{u}$ and tensor $\tens{A}$
that transform consistently under any change of basis:
when the arguments $\vect{u}$ and $\tens{A}$ undergo a rotation, $\vect{F}$ rotates accordingly \citep{Itskov2015}.
The vector field $\vect{F}$ will be finally  determined by requiring that it is linear in $\vect{u}$ and that it solves equation \eqref{A_constr}.

In general, $\vect{F}$ can depend separately on the symmetric and anti-symmetric parts of the velocity gradient \citep{Rivlin1955}
\begin{align}
\tens{S} = \frac{1}{2}\left(\tens{A} + \tens{A}^\top\right), &&
\tens{W} = \frac{1}{2}\left(\tens{A} - \tens{A}^\top\right),
\label{S_and_W}
\end{align}
with $\tens{A}^\top$ denoting the matrix transpose of $\tens{A}$.
Therefore, we consider all the vector functions $\vect{F}(\vect{u},\tens{S},\tens{W})$ constructed through  the velocity and velocity gradients that, due to equation \eqref{phi_u}, are linear in the velocity,
\begin{align}
\vect{F} = \sum_{n=0}^{8} \fvel_{n} (\inv) \tens{B}^n \vect{u} 
\label{F_form}
\end{align}
Here $\tens{B}^n$  are the basis tensors that can be formed through $\tens{S}$ and $\tens{W}$ \citep{Pennisi1987}
\begin{align}
\begin{aligned}[c]
\tens{B}^0 &= \tens{I} \\
\tens{B}^1 &= \tens{S} \\
\tens{B}^2 &= \tens{W}
\end{aligned}
&&
\begin{aligned}[c]
\tens{B}^3  &= \tens{S} \tens{S}\\
\tens{B}^4  &= \tens{S} \tens{W} - \tens{W} \tens{S}\\
\tens{B}^5  &= \tens{S} \tens{W} + \tens{W} \tens{S}
\end{aligned}
&&
\begin{aligned}[c]
\tens{B}^6 &= \tens{W} \tens{W} \\
\tens{B}^7 &= \tens{S} \tens{W} \tens{W} +  \tens{W} \tens{W} \tens{S}  \\
\tens{B}^8 &= \tens{S} \tens{S} \tens{ W} + \tens{W} \tens{S} \tens{S},
\end{aligned}
\label{BT_func}
\end{align}
with $\tens{I}$ denoting the identity matrix and 
standard matrix product implied.
Two additional tensors would be necessary to fix degeneracies of the basis \eqref{BT_func}, which occur when the vorticity is an eigenvector of the strain-rate tensor or the strain-rate has two identical eigenvalues \citep{Rivlin1955}. We ignore that zero-measure configuration of the gradients.
Also, note that the superscript of $\tens B^n$ serves to number the basis tensors rather than indicating powers of the tensor.

The components $\fvel_n$ 
in \eqref{F_form} are functions of the set $\inv$ of independent invariants that can be formed though the velocity gradients \citep{Pennisi1987}
\begin{align}
\inv_1 &= \Tr\big(\tens{S} \tens{S}\big) &&
\inv_3 = \Tr\big(\tens{S} \tens{S} \tens{S}\big) &&
\inv_5 = \Tr\big(\tens{S} \tens{S} \tens{W} \tens{W}\big),
\nonumber\\
\inv_2 &= \Tr\big(\tens{W} \tens{W}\big) &&
\inv_4 = \Tr\big(\tens{S} \tens{W} \tens{W}\big)
\label{inv_func}
\end{align}
with standard matrix product implied.
A sixth invariant would be necessary to fix the orientation/handedness of the vorticity with respect to the strain-rate eigenvectors. We do not consider the sixth invariant as an independent variable since it is determined by the invariants \eqref{inv_func} up to a sign \citep{Lund1992}.

\section{Solution of the equation encoding the homogeneity constraints}

We use the general expression \eqref{F_form} combined with equation \eqref{A_constr} in order to determine the components of $\vect{F}$, $\fvel_n(\inv)$.
This will yield a vector field $\vect{F}$ associated to the homogeneity constraints for the velocity gradient through $\phi(\tens{A}) = \bnabla\bcdot \vect{F}(\vect{u},\tens{A})$ and \eqref{hom}.

Inserting the general expression of $\vect{F}$ \eqref{F_form} into equation \eqref{A_constr} gives, in component notation,
\begin{align}
\left[\frac{\partial \fvel_l}{\partial \inv_k} \frac{\partial \inv_k}{\partial A_{pq}} B^l_{ij}  + 
\frac{\partial \fvel_l}{\partial \inv_k} \frac{\partial \inv_k}{\partial A_{pi}} B^l_{qj} + 
\fvel_l \frac{\partial B^l_{ij}}{\partial A_{pq}}  +
\fvel_l \frac{\partial B^l_{qj}}{\partial A_{pi}}\right]u_j = v_i\delta_{pq} +  v_q\delta_{pi}.
\label{hom_eq_comp}
\end{align}
Here and throughout repeated indices imply summation, unless otherwise specified.
As shown in  Appendix \ref{sec_main_derivatives}, the derivatives of the invariants \eqref{inv_func} can be written as 
\begin{align}
\frac{\partial \inv_k}{\partial A_{pq}} = M_{km}B^m_{pq},
\label{dinv_M}
\end{align}
while  the derivatives of the basis tensors \eqref{BT_func}
can be expressed as
\begin{align}
\frac{\partial B^n_{ij}}{\partial A_{pq}} = \Gamma^{1,n}_{lm} B^l_{ip}B^m_{qj} + \Gamma^{2,n}_{lm} B^l_{iq}B^m_{pj} + \Gamma^{3,n}_{lm} B^l_{ij}B^m_{pq},
\label{dB_Gamma}
\end{align}
with  $0\le l,m,n\le 8$ and $1\le k\le 5$.
The matrix entries $M_{km}$ featured in equation \eqref{dinv_M} are specified in \eqref{def_dinv_M}.
The symbols $\Gamma^{P,n}_{lm}$  in equation \eqref{dB_Gamma} play the role of Christoffel symbols \citep{Grinfeld2013} and their components are listed in  (\ref{Gamma1},\ref{Gamma2},\ref{Gamma3}).
Inserting the derivatives expressions (\ref{dinv_M}, \ref{dB_Gamma}) into equation \eqref{hom_eq_comp}
yields the following independent equations
\begin{subequations}
\begin{align}
\fvel_n\left( \Gamma^{2,n}_{lm} B^l_{iq}B^m_{pj} + \Gamma^{2,n}_{lm} B^l_{qi}B^m_{pj} \right)u_j &= 0, \\
\left[
\frac{\partial \fvel_l}{\partial\inv_k}M_{km}B^l_{ij}B^m_{pq} + \fvel_n\left( \Gamma^{3,n}_{lm} B^l_{ij}B^m_{pq} + \Gamma^{1,n}_{lm} B^l_{qp}B^m_{ij} \right)\right]u_j &= v_i B^0_{pq}. \label{hom_eq_intermediate1}
\end{align}
\label{hom_eq_intermediate}
\end{subequations}
Equations \eqref{hom_eq_intermediate} should hold for all $\vect{u}$ and $\tens{A}$, so that,  separating out the basis tensors we have the following equations for the components
\begin{subequations}
\begin{align}
&\sum_{n=0}^8 \left( \Gamma^{2,n}_{lm} + t(l)\Gamma^{2,n}_{lm}  \right)\fvel_n = 0 && \forall \,  0\le l,m \le 8, \label{eq_fGamma_2}\\
&\sum_{k=1}^5 \frac{\partial \fvel_l}{\partial\inv_k}M_{km} = -\sum_{n=0}^8 \left( \Gamma^{3,n}_{lm}  + t(m)\Gamma^{1,n}_{ml} \right)\fvel_n && \forall \, 0\le l \le 8,\, 1\le m \le 8, \label{eq_fGamma_13}
\end{align}
\end{subequations}
where indices $l,m$ are not contracted, $t(l)=1$ if $\tens{B}^l$ is symmetric and $t(l)=-1$ if $\tens{B}^l$ is anti-symmetric.
In the steps from equation \eqref{hom_eq_intermediate1} to \eqref{eq_fGamma_13}, the components at $m=0$ have been absorbed into the generic right-hand side $v_i B^0_{pq}$ of \eqref{hom_eq_intermediate1}, and therefore equation \eqref{eq_fGamma_13} only concerns components with $m\ge 1$.

The linear system of 81 equations \eqref{eq_fGamma_2} in the nine variables $\fvel_n$, $0\le n\le 8$,  can be solved using symbolic calculus \citep{Sympy2017}. This yields
\begin{align}
\fvel_1=\fvel_2, && \fvel_3=\fvel_5=\fvel_6, && \fvel_4=\fvel_7=\fvel_8=0.
\label{sol_comp_F_intermediate}
\end{align}
Next, equation \eqref{eq_fGamma_13} has a solution only if, for all $l$, the right-hand side is orthogonal to the kernel of $\tens{M}$, but this condition imposes no further constraints on $\fvel_n$.
Finally, with this orthogonality condition ensured, the derivatives $\partial \fvel_l/\partial \inv_k$ are obtained by multiplying  equation \eqref{eq_fGamma_13} by the Moore-Penrose inverse of $\tens{M}$, with components
$M^{-1}_{mk'}$, thus yielding
\begin{subequations}
\begin{align}
\frac{\partial \fvel_0}{\partial\inv_1} &=
\frac{\partial \fvel_0}{\partial\inv_2} = -\frac{1}{2}\fvel_3, \\
\frac{\partial \fvel_0}{\partial\inv_k} &=0 && \forall\, 3\le k\le 5, \\
\frac{\partial \fvel_n}{\partial\inv_k} &= 0 && \forall\, 1\le k\le 5,\; 1\le n\le 8.
\end{align}
\label{sol_comp_F}
\end{subequations}
By solving the straightforward linear system \eqref{sol_comp_F} with the conditions \eqref{sol_comp_F_intermediate}, we obtain the components $\fvel_n$ of $\vect{F}$ that solves equation \eqref{A_constr} and is linear in $\vect{u}$
\begin{align}
\vect{F} = \bar{f}_1\vect{u} + \bar{f}_2\tens{A} \vect{u} + \bar{f}_3\left(\tens{A}^2 - \frac{1}{2}\Tr(\tens{A}^2)\tens{I}\right) \vect{u}, 
\label{hom_sol}
\end{align}
where the $\bar{f}_n$ are arbitrary constants.
The solution \eqref{hom_sol} of equation \eqref{A_constr} encodes all the Betchov constraints since its divergence yields the gradient principal invariants
\begin{align}
\bnabla\bcdot\vect{F} = 
\bar{f}_2 \Tr\big(\tens{A}^2\big) + \bar{f}_3 \Tr\big(\tens{A}^3\big),
\label{Betchov_sol}
\end{align}
thus retrieving equations (\ref{example_first_Betchov}, \ref{example_second_Betchov}) and, by homogeneity, $\avg{\bar{f}_2\Tr(\tens{A}^2) + \bar{f}_3\Tr(\tens{A}^3)}=0$.

The Betchov homogeneity relations, obtained by averaging equation \eqref{Betchov_sol}, are all the possible homogeneity constraints on the single-point statistics of an incompressible and statistically isotropic gradient since they follow from all the independent solutions of equation \eqref{A_constr}.
In other words, no scalar function of the velocity gradient invariants $\inv_1, \dots, \inv_5$ can be written as the divergence of a vector field, other than 
the principal invariants
$\Tr(\tens{A}^2)=\inv_1+\inv_2$ and
$\Tr(\tens{A}^3)=\inv_3+3\inv_4$.

\section{Homogeneity constraints for the velocity gradient and additional quantities}

Equation \eqref{hom_sol} shows that the homogeneity relations for the velocity gradient alone consist only of the two Betchov constraints. However, \eqref{hom_sol} easily generates homogeneity constraints concerning the velocity gradient together with additional variables.
Indeed, the divergence of $\vect{F}$ in \eqref{hom_sol} does not depend on the gradient of the velocity gradient even when $\vect{u}$ in \eqref{hom_sol} is replaced by any scalar, vector or tensor quantity $\tens{q}$ that does not explicitly depend on the velocity gradient itself. This is because $\vect{F}$ solves equation \eqref{A_constr}.
Therefore, for any vector $\vect{q}$, one can construct homogeneity relations for the scalar quantities
\begin{align}
\psi(\tens{A},\bnabla\vect{q})
=
\bnabla\bcdot\left[
\bar{f}_1\vect{q} +
\bar{f}_2\tens{A} \vect{q} + \bar{f}_3\left(\tens{A}^2 - \frac{1}{2}\Tr(\tens{A}^2)\tens{I}\right) \vect{q}
\right],
\label{hom_sol_gen}
\end{align}
where standard matrix-vector product is implied and the left-hand side depends neither on $\vect{q}$ nor on $\bnabla\tens{A}$.
For example, using equation \eqref{hom_sol_gen} with the pressure gradient divided by the fluid density, $\vect{q}=\bnabla P/\rho$, yields the homogeneity relations for the pressure Hessian in incompressible flows, namely
$\avg{A_{ij} \nabla_i\nabla_j P}=0$ and
$\avg{A_{ik}A_{kj}\nabla_i\nabla_j P}=-\rho\avg{(A_{ij}A_{ji})^2}/2$.
Analogously, employing equation \eqref{hom_sol_gen} with the velocity Laplacian, $\vect{q}=\nabla^2 \vect{u}$, gives the homogeneity relations for the Laplacian of a traceless gradient,
$\avg{A_{ij} \nabla^2 A_{ji}}=0$ and
$\avg{A_{ik}A_{kj}\nabla^2 A_{ji}}=0$.
These relations are especially useful for the Lagrangian modelling of velocity gradients \citep{Meneveau2011,Tom2021}.

\section{Conclusions}
We have shown that the Betchov homogeneity relations are all the possible homogeneity constraints for the velocity gradient in incompressible and statistically isotropic turbulence. Our conclusions apply to the single-point statistics of scalar functions of the velocity gradient.
We have shown how our approach to searching for homogeneity relations on the velocity gradient generalizes to constraints involving additional quantities, like the pressure Hessian and the velocity gradient Laplacian.
The presented methodology is also readily applicable to derive homogeneity constraints in less idealized flows (e.g.~axisymmetric flows).
More generally, the outcome of these calculations may help to deal with high-dimensional tensor equations, which are ubiquitous in fluid dynamics.

\backsection[Acknowledgements]{We thank Lukas Bentkamp for insightful comments on the manuscript.}

\appendix
\section{}
\label{sec_main_derivatives}

In this appendix, we compute the derivatives  with respect to the gradient $\tens{A}$ of the basis tensors $\tens{B}^n$ \eqref{BT_func} and invariants $\inv_k$ \eqref{inv_func}.

We start with the derivative of the invariants \eqref{inv_func}, which can be expressed as linear combinations of the basis tensors, as in equation \eqref{dinv_M}.
The matrix $\tens{M}$ featuring the components of the derivatives of the invariants \eqref{inv_func} in the employed basis \eqref{BT_func} is computed by contracting equation \eqref{dinv_M} with the basis tensors,
\begin{align}
M_{km} = Z^{-1}_{ml}B^l_{pq}\frac{\partial \inv_k}{\partial A_{pq}}
=
\left[\begin{matrix}
0 & 2 & 0 & 0 & 0 & 0 & 0 & 0 & 0\\
0 & 0 & -2 & 0 & 0 & 0 & 0 & 0 & 0\\
- \inv_{1} & 0 & 0 & 3 & 0 & 0 & 0 & 0 & 0\\
- \frac{1}{3}\inv_{2} & 0 & 0 & 0 & 0 & -1 & 1 & 0 & 0\\
- \frac{2}{3}\inv_{4} & 0 & 0 & 0 & 0 & 0 & 0 & 1 & -1
\end{matrix}\right],
\label{def_dinv_M}
\end{align}
where  $Z^{lm}=B^l_{pq}B^m_{pq}$ is the metric tensor and $\tens{Z}^{-1}$ denotes its matrix inverse.

Next, we compute the derivatives of the basis tensors, for which we first introduce some notation. The basis tensors \eqref{BT_func} are products of the symmetric and anti-symmetric parts of the velocity gradient \eqref{S_and_W}, which can in turn be expressed through the fourth-order tensors
\begin{align}
Q^{(t)}_{ijpq} = \frac{1}{2}\left(\delta_{ip}\delta_{jq} + t\delta_{iq}\delta_{jp} -\frac{1+t}{3}\delta_{ij}\delta_{pq} \right)
\label{def_Q}
\end{align}
contracted with the  gradient itself, $S_{ij}=Q^{(+1)}_{ijpq}A_{pq}$ and $W_{ij}=Q^{(-1)}_{ijpq}A_{pq}$.
Then, any basis tensor \eqref{BT_func} of degree $d$ consists of a linear combination of the products
\begin{align}
b^{t_1t_2\dots t_d}_{ij} = Q^{(t_1)}_{ik_1p_1q_1} Q^{(t_2)}_{k_1k_2p_2q_2}
\dots Q^{(t_d)}_{k_{d-1}j p_dq_d}
\left(A_{p_1q_1}A_{p_2q_2}
\dots A_{p_dq_d}\right),
\label{def_prod_b}
\end{align}
with $t_l=\pm 1$ and summation over repeated indices, e.g.~$\tens{b}^{+1,-1} = \tens{S}\tens{W}$.
Also, the components of the basis tensors \eqref{BT_func} with respect to the elementary products \eqref{def_prod_b} are constant,
\begin{align}
B^n_{ij} = \sum_{t_1t_2\dots t_d} c^n_{t_1 t_2\dots t_d} b^{t_1t_2\dots t_d}_{ij},
\label{def_B_b}
\end{align}
e.g.~the non-zero components of 
$\tens{B}^4 = \tens{S}\tens{W} - \tens{W}\tens{S}$ are $c^4_{+1,-1}=-c^4_{-1,+1}=1$.

Using equations \eqref{def_prod_b} and \eqref{def_B_b} we can take the derivative of any basis tensor of degree $d$
\begin{align}
\frac{\partial B^{n}_{ij}}{\partial A_{pq}} &= \sum_{t_1t_2\dots t_d} c_{t_1\dots t_d}^n Q^{(t_1)}_{ik_1p_1q_1}Q^{(t_2)}_{k_1k_2p_2q_2} \dots Q^{(t_d)}_{k_{d-1}j p_dq_d} \frac{\partial}{\partial A_{pq}}\left(A_{p_1q_1}A_{p_2q_2}\dots A_{p_dq_d}\right)=\nonumber\\
&=  \sum_{t_1t_2\dots t_d} c_{t_1\dots t_d}^n \sum_{m=1}^d \left[ b^{t_1 t_2 \dots t_{m-1}}_{i k_{m-1}} Q^{(t_m)}_{k_{m-1}k_m pq} b^{t_{m+1}t_{m+2} \dots t_d}_{k_mj}  \right],
\label{dBn}
\end{align}
with $t_l=\pm 1$, $k_0=i$, $k_d=j$ and contraction over repeated indices.
As expected, the derivative of any basis tensor consists of a linear combination of lower-order tensors.
The derivative $\partial(\cdot)/\partial A_{pq}$ is understood as a directional derivative in $\mathbb{R}^{3,3}$ \citep{Itskov2015} and it is traceless due to incompressibility. In the steps to \eqref{dBn} this property is taken into account by the $\tens{Q}^{(t)}$ since $Q^{(t)}_{ijpq}\delta_{pq}=0$ (in fact, $\tens{Q}^{(t)}$ is the tensor derivative of $\tens{b}^t$ with respect to $\tens{A}$).

Inserting the expression of the fourth-order tensors \eqref{def_Q} into equation \eqref{dBn} we can write the derivatives of the basis tensors more explicitly.
For the basis tensors \eqref{BT_func} of degree one (i.e.~for $1\le n \le 2$) we have
\begin{align}
\frac{\partial B^n_{ij}}{\partial A_{pq}} = \frac{1}{2}\sum_{t_1} c_{t_1}^n \left(\delta_{ip}\delta_{qj} + t_1\delta_{iq}\delta_{pj} -\frac{1+t_1}{3}\delta_{ij}\delta_{pq} \right).
\label{dB1}
\end{align}
The derivatives of the basis tensors \eqref{BT_func} of degree two (i.e.~for $3\le n \le 6$) read
\begin{align}
\frac{\partial B^n_{ij}}{\partial A_{pq}} &= 
\frac{1}{2}\sum_{t_1t_2}c_{t_1t_2}^n \Bigg(\delta_{ip}b^{t_2}_{qj} + t_1\delta_{iq}b^{t_2}_{p j} -\frac{1+t_1}{3}b^{t_2}_{i j}\delta_{pq} +
\nonumber\\ &+ 
b^{t_1}_{i p}\delta_{qj} + t_2 b^{t_1}_{i q}\delta_{pj} -\frac{1+t_2}{3}b^{t_1}_{i j}\delta_{pq}\Bigg).
\label{dB2}
\end{align}
Finally, the derivatives of the basis tensors \eqref{BT_func}  of degree three (i.e.~for $7\le n \le 8$) read
\begin{align}
\frac{\partial B^n_{ij}}{\partial A_{pq}} &= 
\frac{1}{2}\sum_{t_1t_2t_3}c_{t_1t_2t_3}^n
\Bigg(
\delta_{ip}b^{t_2 t_3}_{q j} + t_1\delta_{iq}b^{t_2 t_3}_{p j} -\frac{1+t_1}{3}b^{t_2 t_3}_{i j}\delta_{pq} + b^{t_1}_{i p}b^{t_3}_{q j} +  \nonumber\\
&+ t_2b^{t_1}_{i q}b^{t_3}_{p j} -\frac{1+t_2}{3}b^{t_1t_3}_{ij}\delta_{pq} + 
b^{t_1 t_2}_{i p}\delta_{qj} + t_3b^{t_1 t_2}_{i q}\delta_{pj} -\frac{1+t_3}{3}b^{t_1 t_2}_{i j}\delta_{pq}\Bigg).
\label{dB3}
\end{align}
After replacing the symmetric and anti-symmetric parts of $\tens{b}^{t_1}$ and $\tens{b}^{t_1 t_2}$ with the corresponding $\tens{B}^n$, the tensor derivatives (\ref{dB1},\ref{dB2},\ref{dB3}) can be compactly rewritten as contractions of the basis tensors with the Christoffel symbols $\Gamma^{P,n}_{lm}$, as in equation \eqref{dB_Gamma} in the main text.
For the tensor basis \eqref{BT_func} employed here, the non-zero elements of  $\Gamma^{1,n}_{lm}$ are
\begin{align}
\begin{aligned}
\Gamma^{1,1}_{0 0} &= \frac{1}{2} \\
\Gamma^{1,2}_{0 0} &= \frac{1}{2} \\
\Gamma^{1,3}_{0 1} &= \frac{1}{2} \\
\Gamma^{1,3}_{1 0} &= \frac{1}{2} \\
\Gamma^{1,4}_{0 1} &= - \frac{1}{2}
\end{aligned}
&&
\begin{aligned}
\Gamma^{1,4}_{0 2} &= \frac{1}{2} \\
\Gamma^{1,4}_{1 0} &= \frac{1}{2} \\
\Gamma^{1,4}_{2 0} &= - \frac{1}{2} \\
\Gamma^{1,5}_{0 1} &= \frac{1}{2} \\
\Gamma^{1,5}_{0 2} &= \frac{1}{2}
\end{aligned}
&&
\begin{aligned}
\Gamma^{1,5}_{1 0} &= \frac{1}{2} \\
\Gamma^{1,5}_{2 0} &= \frac{1}{2} \\
\Gamma^{1,6}_{0 2} &= \frac{1}{2} \\
\Gamma^{1,6}_{2 0} &= \frac{1}{2} \\
\Gamma^{1,7}_{0 4} &= - \frac{1}{4}
\end{aligned}
&&
\begin{aligned}
\Gamma^{1,7}_{0 5} &= \frac{1}{4} \\
\Gamma^{1,7}_{0 6} &= \frac{1}{2} \\
\Gamma^{1,7}_{1 2} &= \frac{1}{2} \\
\Gamma^{1,7}_{2 1} &= \frac{1}{2} \\
\Gamma^{1,7}_{4 0} &= \frac{1}{4} 
\end{aligned}
&&
\begin{aligned}
\Gamma^{1,7}_{5 0} &= \frac{1}{4} \\
\Gamma^{1,7}_{6 0} &= \frac{1}{2} \\
\Gamma^{1,8}_{0 3} &= \frac{1}{2} \\
\Gamma^{1,8}_{0 4} &= \frac{1}{4} \\
\Gamma^{1,8}_{0 5} &= \frac{1}{4} 
\end{aligned}
&&
\begin{aligned}
\Gamma^{1,8}_{1 2} &= \frac{1}{2} \\
\Gamma^{1,8}_{2 1} &= \frac{1}{2} \\
\Gamma^{1,8}_{3 0} &= \frac{1}{2} \\
\Gamma^{1,8}_{4 0} &= - \frac{1}{4}\\ 
\Gamma^{1,8}_{5 0} &= \frac{1}{4};
\end{aligned}
\label{Gamma1}
\end{align}
the non-zero elements of  $\Gamma^{2,n}_{lm}$ are
\begin{align}
\begin{aligned}
\Gamma^{2,1}_{0 0} &= \frac{1}{2}  \\
\Gamma^{2,2}_{0 0} &= - \frac{1}{2} \\
\Gamma^{2,3}_{0 1} &= \frac{1}{2} \\
\Gamma^{2,3}_{1 0} &= \frac{1}{2} \\
\Gamma^{2,4}_{0 1} &= \frac{1}{2}
\end{aligned}
&&
\begin{aligned}
\Gamma^{2,4}_{0 2} &= \frac{1}{2} \\
\Gamma^{2,4}_{1 0} &= - \frac{1}{2} \\
\Gamma^{2,4}_{2 0} &= - \frac{1}{2} \\
\Gamma^{2,5}_{0 1} &= - \frac{1}{2} \\
\Gamma^{2,5}_{0 2} &= \frac{1}{2}
\end{aligned}
&&
\begin{aligned}
\Gamma^{2,5}_{1 0} &= - \frac{1}{2} \\
\Gamma^{2,5}_{2 0} &= \frac{1}{2} \\
\Gamma^{2,6}_{0 2} &= - \frac{1}{2} \\
\Gamma^{2,6}_{2 0} &= - \frac{1}{2} \\
\Gamma^{2,7}_{0 4} &= \frac{1}{4}
\end{aligned}
&&
\begin{aligned}
\Gamma^{2,7}_{0 5} &= - \frac{1}{4} \\
\Gamma^{2,7}_{0 6} &= \frac{1}{2} \\
\Gamma^{2,7}_{1 2} &= - \frac{1}{2} \\
\Gamma^{2,7}_{2 1} &= - \frac{1}{2} \\
\Gamma^{2,7}_{4 0} &= - \frac{1}{4}
\end{aligned}
&&
\begin{aligned}
\Gamma^{2,7}_{5 0} &= - \frac{1}{4} \\
\Gamma^{2,7}_{6 0} &= \frac{1}{2} \\
\Gamma^{2,8}_{0 3} &= - \frac{1}{2} \\
\Gamma^{2,8}_{0 4} &= \frac{1}{4} \\
\Gamma^{2,8}_{0 5} &= \frac{1}{4}
\end{aligned}
&&
\begin{aligned}
\Gamma^{2,8}_{1 2} &= \frac{1}{2} \\
\Gamma^{2,8}_{2 1} &= \frac{1}{2} \\
\Gamma^{2,8}_{3 0} &= - \frac{1}{2} \\
\Gamma^{2,8}_{4 0} &= - \frac{1}{4} \\
\Gamma^{2,8}_{5 0} &= \frac{1}{4};
\end{aligned}
\label{Gamma2}
\end{align}
the non-zero elements of  $\Gamma^{3,n}_{lm}$ are
\begin{align}
\Gamma^{3,1}_{0 0} = - \frac{1}{3}
&&
\Gamma^{3,3}_{1 0} = - \frac{2}{3}
&&
\Gamma^{3,5}_{2 0} = - \frac{2}{3}
&&
\Gamma^{3,7}_{6 0} = - \frac{2}{3}
&&
\Gamma^{3,8}_{5 0} = - \frac{2}{3}.
\label{Gamma3}
\end{align}
To obtain equation \eqref{dB_Gamma} from the derivatives expressions (\ref{dB1},\ref{dB2},\ref{dB3}), we computed first the Christoffel symbols relative to the independent elements of the set $\{\tens{b}^{t_1},\tens{b}^{t_1t_2},\tens{b}^{t_1t_2t_3}\}$ and then changed the basis to the $\{\tens{B}^n\}$ listed in \eqref{BT_func}.
Indeed, any set of basis tensors can be expressed through the transformation $\widehat{\tens{B}}^{n'}=T^{n'}_n(\inv) \tens{B}^{n}$, with $\tens{T}$ an invertible matrix function of the invariants.
Under this change of basis the Christoffel symbols transform as
\begin{align}
\widehat{\Gamma}^{1,n'}_{l'm'} &= T^{n'}_n \Gamma^{1,n}_{lm} \left(T^{-1}\right)^l_{l'} \left(T^{-1}\right)^m_{m'}, \quad
\widehat{\Gamma}^{2,n'}_{l'm'}  = T^{n'}_n \Gamma^{2,n}_{lm}  \left(T^{-1}\right)^l_{l'} \left(T^{-1}\right)^m_{m'}, \nonumber\\
\widehat{\Gamma}^{3,n'}_{l'm'} &= \left(T^{n'}_n \Gamma^{3,n}_{lm}   + 
\frac{\partial T^{n'}_l}{\partial \inv_k}M_{km}\right)\left(T^{-1}\right)^l_{l'}\left(T^{-1}\right)^m_{m'}.
\end{align}

\bibliographystyle{jfm}
\bibliography{main}

\newcommand{\JFM}{J. Fluid Mech.} \newcommand{\PRF}{Phys. Rev. Fluid}
  \newcommand{\PRE}{Phys. Rev. E} \newcommand{\POF}{Phys. Fluids.}
  \newcommand{\NATCOMM}{Nat. Commun.} \newcommand{\PHYSREP}{Phys. Rep.}
  \newcommand{\PNAS}{Proc. Natl. Acad. Sci. U.S.A.}
  \newcommand{\PRSLONDON}{Proc. Math. Phys. Eng. Sci.} \newcommand{\PRL}{Phys.
  Rev. Lett.} \newcommand{\AREVFLM}{Annu. Rev. Fluid Mech.}
  \newcommand{\PJCOMPSC}{PeerJ Comput. Sci.} \newcommand{\INTJESC}{Int. J. Eng.
  Sci.} \newcommand{\APPLMECHREV}{Appl. Mech. Rev.} \newcommand{\PHYSD}{Physica
  D} \newcommand{\JRATMECHAN}{J. Rat. Mech. Anal.}
\begin{thebibliography}{29}
\expandafter\ifx\csname natexlab\endcsname\relax\def\natexlab#1{#1}\fi
\def\au#1{#1} \def\ed#1{#1} \def\yr#1{#1}\def\at#1{#1}\def\jt#1{\textit{#1}}
  \def\bt#1{#1}\def\bvol#1{\textbf{#1}} \def\vol#1{#1} \def\pg#1{#1}
  \def\publ#1{#1}\def\arxiv#1{#1}\def\org#1{#1}\def\st#1{\textit{#1}}

\bibitem[Alexakis \& Biferale(2018)]{Alexakis2018}
{\sc \au{Alexakis, A.} \& \au{Biferale, L.}} \yr{2018}  \at{Cascades and
  transitions in turbulent flows}.  \jt{\PHYSREP}  \bvol{767--769},
  \pg{1--101}.

\bibitem[Betchov(1956)]{Betchov1956}
{\sc \au{Betchov, R.}} \yr{1956}  \at{An inequality concerning the production
  of vorticity in isotropic turbulence}.  \jt{\JFM}  \bvol{1}~(5),
  \pg{497--504}.

\bibitem[Bragg {\em et~al.\/}(2021)Bragg, Hammond, Dhariwal \& Meng]{Bragg2021}
{\sc \au{Bragg, A.~D}, \au{Hammond, A.~L.}, \au{Dhariwal, R.} \& \au{Meng, H.}}
  \yr{2021}  \at{Hydrodynamic interactions and extreme particle clustering in
  turbulence}.  \jt{arXiv:2104.02758} .

\bibitem[Buaria {\em et~al.\/}(2020)Buaria, Pumir \& Bodenschatz]{Buaria2020}
{\sc \au{Buaria, D.}, \au{Pumir, A.} \& \au{Bodenschatz, E.}} \yr{2020}
  \at{Self-attenuation of extreme events in {N}avier–{S}tokes turbulence}.
  \jt{\NATCOMM}  \bvol{11}.

\bibitem[Carbone \& Bragg(2020)]{Carbone2020}
{\sc \au{Carbone, M.} \& \au{Bragg, A.~D.}} \yr{2020}  \at{Is vortex stretching
  the main cause of the turbulent energy cascade?}  \jt{\JFM}  \bvol{883},
  \pg{R2}.

\bibitem[Davidson(2004)]{Davidson2004}
{\sc \au{Davidson, P.~A.}} \yr{2004} {\em Turbulence: an introduction for
  scientists and engineers\/}.  \publ{Oxford University Press}.

\bibitem[Enciso {\em et~al.\/}(2016)Enciso, Peralta-Salas \&
  de~Lizaur]{Enciso2016}
{\sc \au{Enciso, A.}, \au{Peralta-Salas, D.} \& \au{de~Lizaur, F.~T.}}
  \yr{2016}  \at{Helicity is the only integral invariant of volume-preserving
  transformations}.  \jt{\PNAS}  \bvol{113}~(8),  \pg{2035--2040}.

\bibitem[Grinfeld(2013)]{Grinfeld2013}
{\sc \au{Grinfeld, P.}} \yr{2013} {\em Introduction to Tensor Analysis and the
  Calculus of Moving Surfaces\/}.  \publ{Springer New York}.

\bibitem[Hierro \& Dopazo(2003)]{Hierro2003}
{\sc \au{Hierro, J.} \& \au{Dopazo, C.}} \yr{2003}  \at{Fourth-order
  statistical moments of the velocity gradient tensor in homogeneous, isotropic
  turbulence}.  \jt{\POF}  \bvol{15}~(11),  \pg{3434--3442}.

\bibitem[Hill(1997)]{Hill1997}
{\sc \au{Hill, R.~J.}} \yr{1997}  \at{Applicability of {K}olmogorov's and
  {M}onin's equations of turbulence}.  \jt{\JFM}  \bvol{353},  \pg{67--81}.

\bibitem[Itskov(2015)]{Itskov2015}
{\sc \au{Itskov, M.}} \yr{2015} {\em Tensor Algebra and Tensor Analysis for
  Engineers\/}.  \publ{Switzerland: Springer International Publishing}.

\bibitem[Johnson(2020)]{Johnson2020}
{\sc \au{Johnson, P.~L.}} \yr{2020}  \at{Energy transfer from large to small
  scales in turbulence by multiscale nonlinear strain and vorticity
  interactions}.  \jt{\PRL}  \bvol{124},  \pg{104501}.

\bibitem[Johnson \& Meneveau(2016)]{Johnson2016}
{\sc \au{Johnson, P.~L.} \& \au{Meneveau, C.}} \yr{2016}  \at{A closure for
  lagrangian velocity gradient evolution in turbulence using recent-deformation
  mapping of initially {G}aussian fields}.  \jt{\JFM}  \bvol{804},
  \pg{387--419}.

\bibitem[Leppin \& Wilczek(2020)]{Leppin2020}
{\sc \au{Leppin, L.~A.} \& \au{Wilczek, M.}} \yr{2020}  \at{Capturing velocity
  gradients and particle rotation rates in turbulence}.  \jt{\PRL}  \bvol{125},
   \pg{224501}.

\bibitem[Lund \& Novikov(1992)]{Lund1992}
{\sc \au{Lund, T.S.} \& \au{Novikov, E.~A.}} \yr{1992}  \at{Parameterization of
  subgrid-scale stress by the velocity gradient tensor}.  \jt{Center for
  Turbulence Research (Stanford University and NASA)} .

\bibitem[Majda \& Bertozzi(2001)]{Majda2001}
{\sc \au{Majda, A.~J.} \& \au{Bertozzi, A.~L.}} \yr{2001} {\em Vorticity and
  Incompressible Flow\/}. {\em Cambridge Texts in Applied Mathematics\/} .
  \publ{Cambridge University Press}.

\bibitem[Meneveau(2011)]{Meneveau2011}
{\sc \au{Meneveau, C.}} \yr{2011}  \at{Lagrangian dynamics and models of the
  velocity gradient tensor in turbulent flows}.  \jt{\AREVFLM}  \bvol{43}~(1),
  \pg{219--245}.

\bibitem[Meurer {\em et~al.\/}(2017)Meurer, Smith, Paprocki, \v{C}ert\'{i}k,
  Kirpichev, Rocklin, Kumar, Ivanov, Moore, Singh, Rathnayake, Vig, Granger,
  Muller, Bonazzi, Gupta, Vats, Johansson, Pedregosa, Curry, Terrel,
  Rou\v{c}ka, Saboo, Fernando, Kulal, Cimrman \& Scopatz]{Sympy2017}
{\sc \au{Meurer, A.}, \au{Smith, C.~P.}, \au{Paprocki, M.}, \au{\v{C}ert\'{i}k,
  O.}, \au{Kirpichev, S.~B.}, \au{Rocklin, M.}, \au{Kumar, A.}, \au{Ivanov,
  S.}, \au{Moore, J.~K.}, \au{Singh, S.}, \au{Rathnayake, T.}, \au{Vig, S.},
  \au{Granger, B.~E.}, \au{Muller, R.~P.}, \au{Bonazzi, F.}, \au{Gupta, H.},
  \au{Vats, S.}, \au{Johansson, F.}, \au{Pedregosa, F.}, \au{Curry, M.~J.},
  \au{Terrel, A.~R.}, \au{Rou\v{c}ka, \v{S}.}, \au{Saboo, A.}, \au{Fernando,
  I.}, \au{Kulal, S.}, \au{Cimrman, R.} \& \au{Scopatz, A.}} \yr{2017}
  \at{Sympy: symbolic computing in {P}ython}.  \jt{\PJCOMPSC}  \bvol{3},
  \pg{e103}.

\bibitem[Momenifar {\em et~al.\/}(2021)Momenifar, Diao, Tarokh \&
  Bragg]{Momenifar2021}
{\sc \au{Momenifar, M.}, \au{Diao, E.}, \au{Tarokh, V.} \& \au{Bragg, A.~D.}}
  \yr{2021}  \at{Dimension reduced turbulent flow data from deep vector
  quantizers} ,  \arxiv{arXiv: 2103.01074}.

\bibitem[Pennisi \& Trovato(1987)]{Pennisi1987}
{\sc \au{Pennisi, S.} \& \au{Trovato, M.}} \yr{1987}  \at{On the irreducibility
  of professor {G.F.} {S}mith's representations for isotropic functions}.
  \jt{\INTJESC}  \bvol{25}~(8),  \pg{1059--1065}.

\bibitem[Rivlin \& Ericksen(1955)]{Rivlin1955}
{\sc \au{Rivlin, R.~S.} \& \au{Ericksen, J.~L.}} \yr{1955}
  \at{Stress-deformation relations for isotropic materials}.  \jt{\JRATMECHAN}
  \bvol{4},  \pg{323--425}.

\bibitem[Serre(1984)]{Serre1984}
{\sc \au{Serre, D.}} \yr{1984}  \at{Les invariants du premier ordre de
  l'equation d'{E}uler en dimension trois}.  \jt{\PHYSD}  \bvol{13}~(1),
  \pg{105--136}.

\bibitem[Siggia(1981)]{Siggia1981}
{\sc \au{Siggia, E.~D.}} \yr{1981}  \at{Invariants for the one‐point
  vorticity and strain rate correlation functions}.  \jt{\POF}  \bvol{24}~(11),
   \pg{1934--1936}.

\bibitem[Tian {\em et~al.\/}(2021)Tian, Livescu \& Chertkov]{Tian2021}
{\sc \au{Tian, Y.}, \au{Livescu, D.} \& \au{Chertkov, M.}} \yr{2021}
  \at{Physics-informed machine learning of the {L}agrangian dynamics of
  velocity gradient tensor}.  \jt{\PRF}  \bvol{6},  \pg{094607}.

\bibitem[Tom {\em et~al.\/}(2021)Tom, Carbone \& Bragg]{Tom2021}
{\sc \au{Tom, J.}, \au{Carbone, M.} \& \au{Bragg, A.~D.}} \yr{2021}
  \at{Exploring the turbulent velocity gradients at different scales from the
  perspective of the strain-rate eigenframe}.  \jt{\JFM}  \bvol{910},
  \pg{A24}.

\bibitem[Townsend \& Taylor(1951)]{Townsend1951}
{\sc \au{Townsend, A.~A.} \& \au{Taylor, G.~I.}} \yr{1951}  \at{On the
  fine-scale structure of turbulence}.  \jt{\PRSLONDON}  \bvol{208}~(1095),
  \pg{534--542}.

\bibitem[Tsinober(2009)]{Tsinober2009}
{\sc \au{Tsinober, A.}} \yr{2009} {\em An Informal Conceptual Introduction to
  Turbulence\/}.  \publ{Springer Netherlands}.

\bibitem[Weyl(1946)]{Weyl1946}
{\sc \au{Weyl, H.}} \yr{1946} {\em The Classical Groups: Their Invariants and
  Representations\/}. {\em Princeton Landmarks in Mathematics and Physics\/}
  Nr. 1,Teil 1.  \publ{Princeton University Press}.

\bibitem[Zheng(1994)]{Zheng1994}
{\sc \au{Zheng, Q.-S.}} \yr{1994}  \at{{Theory of Representations for Tensor
  Functions—A Unified Invariant Approach to Constitutive Equations}}.
  \jt{\APPLMECHREV}  \bvol{47}~(11),  \pg{545--587}.

\end{thebibliography}

\end{document}